\documentclass[11pt]{article}

\usepackage[preprint]{nlp4MusA}

\usepackage{times}
\usepackage{latexsym}
\usepackage[T1]{fontenc}
\usepackage[utf8]{inputenc}
\usepackage{microtype}
\usepackage{inconsolata}
\usepackage{graphicx}
\usepackage{booktabs}
\usepackage{url}
\usepackage{subcaption}
\usepackage{amssymb}
\usepackage{makecell}

\title{LabelBuddy: An Open Source Music and Audio Language Annotation Tagging Tool Using AI Assistance}

\author{
  \textbf{Ioannis Prokopiou\textsuperscript{1,2}},
  \textbf{Ioannis Sina\textsuperscript{3}},
  \textbf{Agisilaos Kounelis\textsuperscript{3}},
  \textbf{Pantelis Vikatos\textsuperscript{2}},
  \textbf{Themos Stafylakis\textsuperscript{1,4}}
\\
\\
  \textsuperscript{1}Athens University of Economics and Business, \quad
  \textsuperscript{2}Orfium, \quad
  \textsuperscript{3}University of Patras, \quad
  \textsuperscript{4}Archimedes/Athena R.C., 
\\
  \small{
    \texttt{gian.prokopiou@aueb.gr}, \texttt{sinaioannis@gmail.com}
  } \\
  \small{
    \texttt{agis@ceid.upatras.gr}, \texttt{pantelis@orfium.com}, \texttt{tstafylakis@aueb.gr}
  }
}

\begin{document} 
\maketitle 

\begin{abstract} 
The advancement of Machine learning (ML), Large Audio Language Models (LALMs), and autonomous AI agents in Music Information Retrieval (MIR) necessitates a shift from static tagging to rich, human-aligned representation learning. However, the scarcity of open-source infrastructure capable of capturing the subjective nuances of audio annotation remains a critical bottleneck. This paper introduces \textbf{LabelBuddy}, an open-source collaborative auto-tagging audio annotation tool designed to bridge the gap between human intent and machine understanding. Unlike static tools, it decouples the interface from inference via containerized backends, allowing users to plug in custom models for AI-assisted pre-annotation. We describe the system architecture, which supports multi-user consensus, containerized model isolation, and a roadmap for extending agents and LALMs. Code available at \url{https://github.com/GiannisProkopiou/gsoc2022-Label-buddy}.
\end{abstract} 


\section{Introduction} 
\label{sec:intro}


The quality of the datasets used to train AI models constitutes a significant factor in accuracy, reliability, and generalization~\cite{picard2020ensuring}. Despite standardization efforts through public repositories like Zenodo\footnote{https://zenodo.org/} and data lakes~\cite{espinal_xavier_2022_6463482}, or benchmarks like DCASE\footnote{https://dcase.community/} and MIREX\footnote{https://www.music-ir.org/mirex}, available resources are often task-specific. In addition, the creation of specialized datasets can be a challenging procedure, complicated, time-consuming, and is often laborious and supported by human review~\cite{voigtlaender2021reducing}. Building on the tradition of cross-disciplinary impact, the intersection of Natural Language Processing (NLP) with music and audio presents a frontier where language and sound converge. Most audio content contains an inherent linguistic dimension, yet creating datasets to capture these multimodal synergies remains a laborious bottleneck.

The domain of Music Information Retrieval (MIR) is currently undergoing a transition from discriminative paradigms characterized by static tag classification to generative and reasoning-based approaches. The rise of Large Audio-Language Models (LALMs) such as Music Flamingo~\cite{ghosh2025music}, Qwen-Audio \cite{chu2023qwen}, and Audio Flamingo 3~\cite{goel2025audio} has introduced new capabilities for "chain-of-thought" reasoning and conversational audio understanding. However, the efficacy of these models is heavily dependent on the quality of alignment with human intent. Recent surveys indicate that objective metrics often fail to capture aesthetic nuance, necessitating a pivot toward Reinforcement Learning from Human Feedback (RLHF) and rigorous subjective evaluation methodologies~\cite{kader2025survey}.

Current workflows are often fragmented, and users resort to disjointed workflows, separating data curation from the critical phase of manual subjective evaluation (e.g., MUSHRA, GoListen, or pairwise preference testing) \cite{schoeffler2018webmushra, barry2021go}. Users use waveform-based tools for segmentation \cite{grover2020audino}, separate platforms for text handling, and distinct software for subjective evaluation (e.g., WebMUSHRA \cite{schoeffler2018webmushra}). This separation hinders the development of efficient \textit{Human-in-the-Loop} (HITL) pipelines, where the uncertainty of the model's output should drive data acquisition. Furthermore, the "crisis of metrics" in generative music, where objective scores like FAD fail to correlate with human perception \cite{kader2025survey} demands tools that can seamlessly transition from annotation to subjective preference ranking.

To address this, we present \textbf{LabelBuddy}, an open-source collaborative auto-tagging audio annotation tool equipped with AI assistance with:
\begin{enumerate}
    \item \textbf{Decoupled AI-Assistance:} An isolated containerized architecture injects model predictions via declarative YAML files. We provide pre-trained models like YOHO~\cite{venkatesh2022you}, musicnn~\cite{pons2019musicnn}, PANNs~\cite{kong2020panns}, and LALMs like Music Flamingo~\cite{ghosh2025music} for AI-assisted pre-annotation tags to shift user effort from creation to verification, while approved labels can be used to fine-tune the models.
    \item \textbf{Collaborative Consensus:} Native support for multi-user roles (manager, annotator, reviewer) to ensure ground-truth reliability.
    \item \textbf{Hybrid Workflow Support:} An architecture designed to support both region-based tagging and subjective preference aggregation.
\end{enumerate}

\section{Related Work}
\label{sec:related}


This section reviews annotation platforms, HITL workflows for LALMs, and infrastructure for subjective evaluation and RLHF.

\paragraph{Annotation Platforms \& Domain Specificity.}
The landscape of data curation significantly varies by modality. For text, tools like BRAT~\cite{stenetorp2012brat}, Paladin~\cite{nghiem2021paladin}, and PubAnnotation~\cite{kim2012pubannotation} facilitate linguistic tagging, while specialized frameworks like CAT~\cite{bartalesi2012cat} and MDSWriter~\cite{meyer2016mdswriter} handle semantic efficiency and summarization respectively. Active learning strategies have been explored in text labeling (e.g., ActiveAnno~\cite{wiechmann2021activeanno}, APLenty~\cite{nghiem2018aplenty}). In the visual domain, tools like VIA~\cite{dutta2019via} and Annotation Web~\cite{smistad2021annotation} show the need for domain-specific interfaces.


In the audio domain, tools like Audino~\cite{grover2020audino} and BAT~\cite{melendez2017bat} excel at temporal tasks like Sound Event Detection (SED) and salience, while library-based solutions like Aubio~\cite{paul_brossier_2019_2578765} facilitate feature extraction. Others, like Gecko~\cite{levy2019gecko}, focus on voice segmentation. However, these tools generally lack the decoupled AI architecture required for modern reasoning model use. Conversely, general-purpose HITL platforms like Label Studio~\cite{label_studio} and Prodigy~\cite{montani2018prodigy} offer robust backends but often restrict collaborative features such as reviewer roles and consensus metrics to paid enterprise tiers. Furthermore, they lack native support for musical structures (e.g., bars, beats) found in specialized audio tools~\cite{cartwright2017seeing}. A comparison between LabelBuddy and other existing tools is shown in Table \ref{tab:comparison}.

\paragraph{LALMs \& HITL Workflows.}
The state-of-the-art has shifted from fixed-vocabulary auto-taggers  to Large Audio-Language Models (LALMs) such as Audio Flamingo 3~\cite{goel2025audio} and Qwen-Audio \cite{chu2023qwen}. These utilize unified encoders for "chain-of-thought" reasoning. To align them, we adopt a "Single-Iteration" HITL approach. Recent studies in video annotation~\cite{gutierrez2025video} demonstrate that simple model-assisted pre-annotation reduces time-on-task without degrading quality, a philosophy we extend to audio similarly to NEAL~\cite{gibbons2023neal}.

\paragraph{Subjective Evaluation \& RLHF.}
A critical bottleneck in generative music is the "crisis of metrics," where scores like Fréchet Audio Distance (FAD) fail to correlate with human perception~\cite{kader2025survey, gui2023adapting}. Consequently, the field is pivoting towards RLHF~\cite{cideron2024musicrl, liu2025musiceval}. Currently, evaluation is decoupled from annotation, relying on standalone tools like WebMUSHRA~\cite{schoeffler2018webmushra}. LabelBuddy aims to unify this by integrating pairwise preference aggregation methods, such as Bayesian Bradley–Terry (BBQ)~\cite{aczel2025bbq}.

\begin{table}[t]
\centering
\caption{Comparison of LabelBuddy with existing tools. 
}
\label{tab:comparison}
\resizebox{\columnwidth}{!}{%
\begin{tabular}{lccccc}
\toprule
\textbf{Tool} & 
\textbf{\makecell{Audio\\Specific}} & 
\textbf{\makecell{Decoupled\\AI-Assist}} & 
\textbf{\makecell{Open\\Source}} & 
\textbf{\makecell{Collaboratory\\Consensus}} \\
\midrule
Audino & \checkmark & - & \checkmark & - \\
BAT & \checkmark & -  & \checkmark & - \\
Aubio & \checkmark & - & \checkmark & - \\
Gecko & \checkmark & - & \checkmark & - \\
Prodigy & - & \checkmark  & - & - \\
Label Studio (CE) & - & \checkmark & \checkmark & - \\
\midrule
\textbf{LabelBuddy} & \textbf{\checkmark} & \textbf{\checkmark} & \textbf{\checkmark} & \textbf{\checkmark} \\
\bottomrule
\end{tabular}%
}
\end{table}


\section{System Architecture}
\label{sec:backbone}

\begin{figure}[t]
  

    \includegraphics[width=\linewidth]{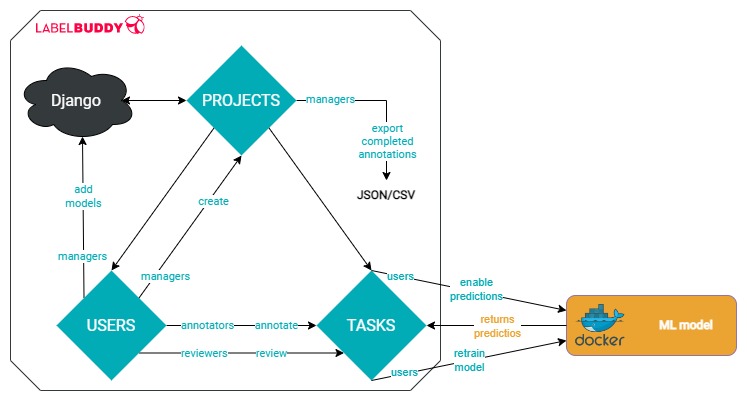}
    \caption{\textbf{System Architecture Overview}: The architecture decouples the Django web server from Dockerized ML inference.}
    \label{fig:backend}

  \label{fig:sys_arch}
\end{figure}

LabelBuddy addresses the "coupling problem" in annotation tools where interfaces are hard-coded to specific model backends via a modular, containerized architecture. As illustrated in Figure \ref{fig:sys_arch}, the system decouples the lightweight user interaction layer (Django) from the compute-intensive inference layer (Docker).

\subsection{Backend \& Data Model}
The core application is built on Django, utilizing a relational database to manage the three primary entities: \textit{Projects}, \textit{Users}, and \textit{Tasks}.
\begin{itemize}
    \item \textbf{RBAC \& Privacy:} To prevent data leakage, the system implements Role-Based Access Control (RBAC). Managers have full oversight, while \textit{Annotators} and \textit{Reviewers} are restricted to their assigned task queues.
    \item \textbf{Data Serialization:} Annotations are stored as JSON objects that contain temporal boundaries and ontology tags. Managers can export consensus data to CSV/JSON formats for direct integration with ML training pipelines.
\end{itemize}

\subsection{Containerized Inference Engine}
 Managers define models via a YAML configuration file, specifying the Docker image, input/output schema, and resource constraints. When AI assistance is requested, the backend communicates with the model container via a RESTful Flask API. This design ensures Sandboxing (models run in isolated environments) and Scalability (inference can be deployed on remote cloud nodes like AWS/Azure).

\begin{figure*}[t]
  

    \centering
    \includegraphics[width=0.70\linewidth]{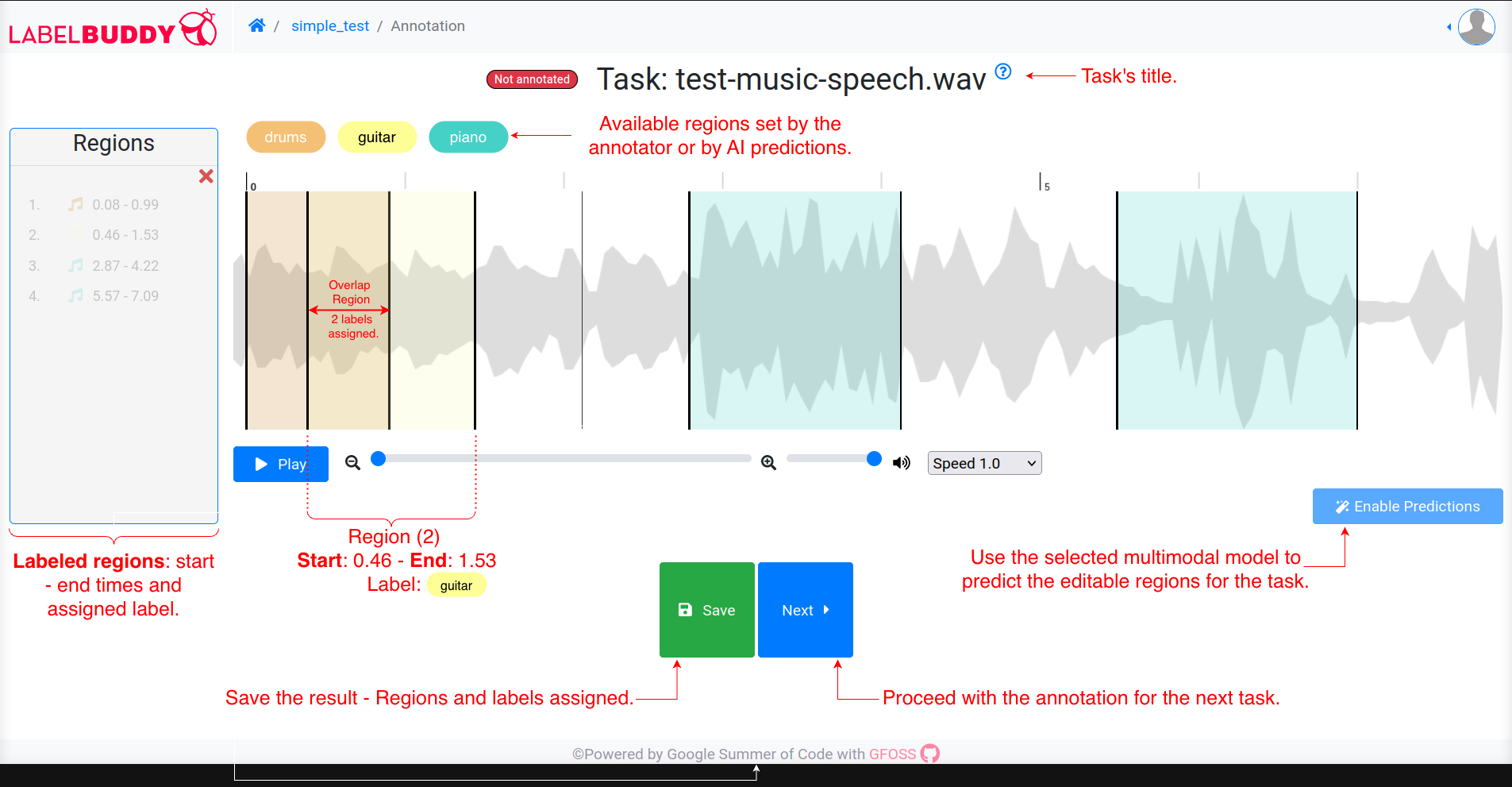}
    \caption{The \textbf{annotation interface} displaying AI-generated predictions as editable waveform regions.}
    \label{fig:annotation}

  \label{fig:full_system}
\end{figure*}

\section{Workflow \& Interface}
\label{sec:workflow}

The platform supports a comprehensive "Human-in-the-Loop" (HITL) lifecycle.

\subsection{Project Setup \& Task Management}
The workflow begins in the \textbf{Dashboard}, where managers create projects and assign user roles.
\begin{itemize}
    \item \textbf{Model Integration:} In the \textbf{Model Page}, managers upload YAML configuration files to attach inference containers. This interface exposes advanced controls: monitoring training loss/accuracy, downloading weight files, and triggering fine-tuning jobs (specifying epochs and learning rates) using the project's validated data.
    \item \textbf{Task Ingestion:} Managers upload audio (WAV/MP3) via the \textbf{Project Page}, which can be distributed to annotators via a shared pool or disjoint assignment strategies.
\end{itemize}

\subsection{The Annotation Loop}
The core labeling workflow utilizes \texttt{wavesurfer.js} for responsive waveform visualization.
\begin{enumerate}
    \item \textbf{AI-Assisted Pre-Annotation:} Annotators trigger "On-Demand Prediction," which serializes the audio to the active Docker container. The system renders the returned predictions as editable regions (Fig. \ref{fig:full_system}), shifting the human task from \textit{creation} to \textit{verification}.
    \item \textbf{Review \& Consensus:} Completed tasks enter the \textbf{Review Interface}, where reviewers can play back specific regions and approve or reject annotations with feedback. This Quality Assurance (QA) loop is essential for creating high-fidelity datasets for generative alignment.
\end{enumerate}

\section{Case Study: NLP  Music Tagging}
\label{sec:usecase}

To demonstrate LabelBuddy's utility, we present a reference workflow for creating a Music Captioning Dataset, a task requiring the alignment of audio signals with rich natural language descriptions.

\paragraph{Model Integration} 
The project manager defines a Docker container that wraps a multimodal model, such as a Music Flamingo checkpoint. The YAML configuration maps the model's text output to a LabelBuddy \texttt{Annotation} region:

\begin{small}
\begin{verbatim}
image: "my-repo/music-flamingo:v1"
input_schema: { "audio": "wav" }
output_schema: 
  - { "type": "text", "label": "Caption" }
resources: { "gpu": "true" }
\end{verbatim}
\end{small}

\paragraph{The "Human-Verify" Loop.} 
Annotators are presented with a queue of raw audio tracks. Instead of writing descriptions from scratch (which is cognitively demanding), they trigger the \textbf{Pre-Annotate} function. The backend container processes the audio and returns a candidate caption: \textit{"A lo-fi hip-hop track with a slow tempo and vinyl crackle."}

\paragraph{Correction \& Consensus.} 
The annotator corrects specific hallucinations (e.g., changing "vinyl crackle" to "rain sounds") and adjusts timestamp boundaries. If multiple annotators process the same track, the Reviewer Interface highlights semantic disagreements in the captions, allowing for the creation of robust, consensus-based ground truth.

\paragraph{Multimodal Export.} 
The finalized dataset is exported as a JSONL or CSV file containing aligned \texttt{(audio\_path, text\_caption)} pairs, ready for immediate use in fine-tuning downstream audio-to-text generation models.

\section{Discussion \& Future Roadmap}
\label{sec:future}

The current release of LabelBuddy solves the immediate infrastructure challenge: decoupling the annotation frontend from rapidly evolving model backends. As the field transitions towards LALMs and autonomous agents, our future roadmap is aligned with bridging the gap between human intent and machine representation:

\textbf{Agentic Reasoning.}
While traditional active learning relies on uncertainty sampling, the rise of LALMs requires a shift toward \textit{conversational} assistance. We are extending the backend API to support more reasoning-capable models such as Qwen-Audio \cite{chu2023qwen}. Future versions will allow annotators to query the model and receive "Chain-of-Thought" justifications, transforming the workflow from simple tag verification to collaborative reasoning. This aligns with recent findings that interactive reasoning reduces hallucination in complex annotation tasks.

\textbf{Integrated Subjective Evaluation (RLHF).}
Acknowledging the "crisis of metrics" where FAD scores fail to capture aesthetic quality \cite{kader2025survey}, LabelBuddy aims to evolve into a workbench for RLHF. We aim to implement a native "Pairwise Preference" interface (Dataset A vs. Dataset B) directly in the review loop. To handle noisy human raters, the backend will integrate {Bayesian Bradley–Terry (BBQ) models \cite{aczel2025bbq}, providing robust preference aggregation to align generative models \cite{cideron2024musicrl}.

\textbf{Enhancing Perceptual Validity.}
To counteract the tendency of models to rely on text priors rather than audio content a flaw highlighted by the RUListening benchmark \cite{zang2025are}, we plan to introduce timestamp-required QA templates. These will force both models and human annotators to ground every semantic claim in specific spectral regions, ensuring that future datasets drive genuine auditory perception rather than text-only reasoning.

\textbf{Evaluation Plan.} To validate utility, we propose a pilot study on DCASE 2024 data measuring: (a) time reduction vs. \textit{de novo} labeling, (b) inter-annotator agreement (Fleiss' Kappa), and (c) downstream PSDS gains for baseline SED models trained on LabelBuddy-curated data.

\section{Conclusion}
\label{sec:conclusion}

LabelBuddy serves as critical infrastructure for exploring the multimodal synergies between language and audio. By decoupling the interface from inference, it empowers the community to curate the rich, linguistically-grounded datasets required for modern NLP-driven music understanding. Whether for standard tagging or the emerging demands of RLHF, LabelBuddy offers an open, scalable workbench to deepen the connection between human perception and machine representation on audio.

\section{Ethics Statement} The development of AI-assisted annotation tools raises concerns regarding labor displacement and bias. LabelBuddy is designed to augment, not replace, human expertise, keeping the human in the loop for critical judgments. Furthermore, by facilitating the creation of open datasets, we aim to democratize access to high-quality training data, countering the centralization of resources in large tech corporations. We ensure that all integrated models are used in compliance with their research licenses.

\section*{Acknowledgments}
This work was supported by Google Summer of Code (GSoC) and the Open Technologies Alliance (GFOSS) and was funded by the European Union’s Horizon Europe research and innovation programme under the AIXPERT project (Grant Agreement No. 101214389), which aims to develop an agentic, multi-layered, GenAI-powered framework for creating explainable, accountable and transparent AI systems.

\bibliography{nlp4MusA}

\end{document}